\documentclass[aps,prc,superscriptaddress,showpacs,floatfix,twocolumn]{revtex4-1}

\usepackage{lineno}
\usepackage{graphicx}
\usepackage{color}

\newcommand{\mean}[1]{$\left\langle #1 \right\rangle$} 
 
\newcommand{\pt}{$p_T$}

\newcommand{\auau}{\mbox{Au+Au} }

\newcommand{\degree}[1]{$#1^{\circ}$}

\newcommand{\gevc}{GeV/{\it c}}

\newcommand{\snn}{$\sqrt{s_{_{\rm NN}}}$}

\begin{document}
\title{Rapidity dependence of  deuteron production in central \auau collisions at \snn~ = 200 GeV }
\newcommand{\bnl}{~Brookhaven National Laboratory, Upton, NY 11973-5000, U.S.}
\newcommand{\krakow}{~Smoluchowski Inst. of Physics, Jagiellonian University, Krakow, Poland}
 \newcommand{\newyork}{~New York University, New York, NY 10003}
 \newcommand{\nbi}{~Niels Bohr Institute, Blegdamsvej 17, University of Copenhagen, Copenhagen, Denmark}
 \newcommand{\texas}{~Texas A$\&$M University, College Station, TX 17843}
 \newcommand{\bergen}{~University of Bergen, Department of Physics and Technology, Bergen, Norway}
 \newcommand{\bucharest}{~University of Bucharest, Bucharest, Romania}
 \newcommand{\kansas}{~University of Kansas, Lawrence, KS 66045}
 \newcommand{\oslo}{~University of Oslo, Department of Physics, Oslo, Norway}
\newcommand{\spacescience}{~Institute for Space Sciences, Bucharest, Romania}
 \newcommand{\ires}{~Institute Pluridisciplinaire Hubert Curien CRNS-IN2P3 et
  Universit{\'e} de  Strasbourg, Strasbourg, France}

\affiliation{\bnl}
\affiliation{\ires}
\affiliation{\newyork}
\affiliation{\krakow}
\affiliation{\nbi}
\affiliation{\spacescience}
\affiliation{\texas}
\affiliation{\bergen}
\affiliation{\bucharest}
\affiliation{\kansas}
\affiliation{\oslo}

\author{I. Arsene}\altaffiliation[Present Address: ]{ExtreMe Matter Institute EMMI, GSI,Darmstadt, Germany}\affiliation{\oslo}
\author{I.~G.~Bearden} \affiliation{\nbi}
\author{D.~Beavis} \affiliation{\bnl}
\author{S.~Bekele}\altaffiliation[Present address: ]{Dept. of Physics, Tennessee Tech University,Cookeville, Tennessee}\affiliation{\kansas}
\author{C.~Besliu} \affiliation{\bucharest}
\author{B.~Budick}\affiliation{\newyork}
\author{H.~B{\o}ggild} \affiliation{\nbi}
\author{C.~Chasman} \affiliation{\bnl}
\author{C.~H.~Christensen}\affiliation{\nbi}
\author{P.~Christiansen}\altaffiliation[Present Address: ]{Div. of Experimental
High-Energy Physics, Lund University, Lund, Sweden}\affiliation{\nbi}
\author{H.~H.~Dalsgaard} \affiliation{\nbi}
\author{R.~Debbe} \affiliation{\bnl} 
\author{J.~J.~Gaardh{\o}je} \affiliation{\nbi}
\author{K.~Hagel}\affiliation{\texas} 
\author{H.~Ito} \affiliation{\bnl}
\author{A.~Jipa} \affiliation{\bucharest}
\author{E.~B.~Johnson}\altaffiliation[Present address: ]{Radiation Monitoring Devices,  Cambridge, MA, USA}\affiliation{\kansas}
\author{C.~E.~J{\o}rgensen}\altaffiliation[Present address: ]{Ris{\o} National Laboratory, Denmark}\affiliation{\nbi}
\author{R.~Karabowicz} \affiliation{\krakow}
\author{N.~Katrynska} \affiliation{\krakow}
\author{E.~J.~Kim}\altaffiliation[Present address: ]{Division of Science Education, Chonbuk National University, Jeonju, 561-756, Korea}\affiliation{\kansas}
\author{T.~M.~Larsen} \affiliation{\nbi}
\author{J.~H.~Lee} \affiliation{\bnl} 
\author{G.~L{\o}vh{\o}iden} \affiliation{\oslo} 
\author{Z.~Majka} \affiliation{\krakow} 
\author{M.~Murray}\affiliation{\kansas} 
\author{J.~Natowitz}\affiliation{\texas} 
\author{B.~S.~Nielsen} \affiliation{\nbi} 
\author{C.~Nygaard}\affiliation{\nbi}
\author{D.~Ouerdane} \affiliation{\nbi}
\author{D.~Pal}\affiliation{\kansas} 
\author{A.~Qviller} \affiliation{\oslo}
\author{F.~Rami}\affiliation{\ires}
\author{C.~Ristea} \affiliation{\nbi}
\author{O.~Ristea} \affiliation{\bucharest}
\author{D.~R{\"o}hrich}\affiliation{\bergen}
\author{S.~J.~Sanders}\affiliation{\kansas}
\author{P.~Staszel} \affiliation{\krakow}
\author{T.~S.~Tveter} \affiliation{\oslo}
\author{F.~Videb{\ae}k}\altaffiliation[]{Spokesperson}\affiliation{\bnl}
\author{R.~Wada}\affiliation{\texas}
\author{H.~Yang}\altaffiliation[Present address: ]{University of Heidelberg, Heidelberg, Germany}\affiliation{\bergen}
\author{Z.~Yin}
\author{S.Zgura}\affiliation{\spacescience}

\collaboration{The BRAHMS Collaboration}\noaffiliation
\date{\today}

\begin{abstract}
We have measured the distributions  
of protons and deuterons  produced in the 20\% most central Au+Au collisions at RHIC (\snn~ = 200 GeV)
over a very wide range of transverse and longitudinal momentum.  Near midrapidity we have also measured the distribution of antiprotons and antideuterons.  We present our results in the context of coalescence models. In particular we extract the ``homogeneity volume'' and the average phase-space density for protons and antiprotons. 
Near central rapidity  the coalescence parameter $B_2(p_T)$ and the space-averaged phase-space density $\langle f \rangle (p_T)$  are very similar for both protons and antiprotons. 
For protons we  see little variation of either  $B_2(p_T)$ or the space-averaged phase-space density as the rapidity increases from 0 to 3. 
However, these quantities depend strongly on $p_T$ at all rapidities.   
These results are   in  contrast to data from lower energy collisions where the proton and antiproton  
phase-space densities are different at $y$=0 and both 
$B_2$ and  \mean{f} depend strongly on rapidity.
\end{abstract}

\maketitle

\section{Introduction}
 Deuterons  detected in heavy ion collisions are conventionally thought to be produced  predominantly via a process called coalescence. 
Protons and neutrons that are close enough in phase-space, (i.e. in position and momentum space) ``coalesce'' to form deuterons.
The surrounding medium created in the A+A collisions enables this $"2\rightarrow$1" process to proceed while
conserving 
energy and momentum.
There is a wide range of evidence to suggest that a dense system of strongly interacting partons is created in heavy ion collisions at Relativistic Heavy Ion Collider (RHIC) energies \cite{WhitePapers}. 
In the hot and dense system produced in high-energy ion collisions, the coalescence of nucleons into deuterons 
cannot even begin 
before the  
the partons have frozen out into hadrons. 
Even then, the low binding energy of 
deuterons ($\sim2.2$ MeV),
ensures that only deuterons produced late will survive to reach the detectors. Thus our sample is dominated by deuterons produced 
 close to  thermal freeze-out, where the nuclear density is low \cite{Ioffe, Leupold:1993ms}. 

Early coalescence models assumed that the phase-space density of 
 clusters is  proportional to the product of the phase-space densities
of individual nucleons that coalesced into them 
and that the momentum 
of the deuteron is trivially the sum of the nucleon momenta 
\cite{BUTLER, KAPUSTA, MEKJIAN}. More recent coalescence models
add insight into the nature of the phenomenon, but still relate cluster production to 
the product of phase-space
densities \cite{Csernai}.  For deuterons this is written as:
\begin{equation}\label{eq:coal-eq}
 E_d \cdot \frac{d^{3}N_d}{dp_d^{3}} = B_{2}({\bf p}) (E_{p} \cdot \frac{d^{3}N_{p}}{dp_{p}^{3}})^{2},
\end{equation}
where  ${\bf p}$ is the momentum of the proton and $B_2$ is defined as the coefficient linking the square of the nucleon distribution to that of the proton distribution.

Ideally for a coalescence study one would measure the spectra of both protons and neutrons. At \snn~= 4.9 GeV the 
$n/p$ ratio  
measured in the 10\% most central Au on Pb collisions has a 
value of $1.19\pm.08$ and is independent of $m_T$ 
in a wide range of rapidity (1.6 to 2.4) \cite{E864Neut}, whereas the same ratio for the incident nuclei 
is equal to 1.52.
(Measurements by other experiments imply that most of the isospin lost by the neutrons goes into an excess of 
$\pi^-$ over $\pi^+$ \cite{E86696,E87796})
At higher energies we expect $n/p$ to be close to 1 over a wide rapidity range.

In thermodynamic models that assume thermalized distributions of nucleons, $B_{2}$  carries information about 
the effective volume (in coordinate space) of the nucleons when they coalescence.  
This is also known as
 the homogeneity volume, 
and is defined as that volume over which nucleons, are close enough in momentum that they can coalesce
  \cite{BUTLER, KAPUSTA, MEKJIAN}. 
 At RHIC the homogeneity volume of the emitting source has been studied in great detail at central rapidity but few measurements are available in the forward region. Consequently the deuteron  coalescence  analysis gives BRAHMS a unique opportunity to study the volume of the proton source over a wide range of rapidity.
 
The paper is organized as follows. This introduction continues with a short
review of the homogeneity volume and the 
phase-space density averaged over coordinate space.
Sec. II
details the detector setup, the selection of the central events used for this 
analysis, the tracking algorithm, particle identification and the feed down corrections
applied to the proton and antiproton spectra. In Sec. III we describe 
the fully corrected spectra and the inferred values of $B_{2}$ and average phase-space density as 
functions of transverse momentum and rapidity. Finally, we summarize our results and the physics that 
we extract from them.

\subsection{$B_{2}$ and the Homogeneity Volume}
The link between
$B_{2}$ and the homogeneity volume can be seen in coalescence models that 
use the 
so-called
 sudden approximation technique to write the phase-space density  
of deuterons 
as an overlap of the nucleon wave functions with the deuteron wave function. 
The volume of the nucleon system is introduced through the
normalization of their wave functions \cite{BKJennings}. 
One can make a reasonable approximation of the deuteron's wave function by assuming it is a Gaussian of width $\delta = 2.8$ fm. If we further assume 
 that the region where coalescence occurs also has a Gaussian spatial profile with width $R_G$, one can write \cite{Llope}: 
\begin{equation}
(R_G^2({\bf p}) + \frac{\delta^2}{2})^{3/2} = \frac{3}{2} \cdot 
\frac{{\pi}^{3/2}{\hbar}^3}{B_2({\bf p}) \cdot m_p c^2} ,
\label{eq:Rg}
\end{equation}
where $m_p$ denotes the proton mass.   
The left-hand side of this equation just represents the convolution of the proton source size and the Gaussian representation of the deuteron wave-function. For small systems $\delta$ dominates over $R_G$ and the measurements have only little sensitivity to the proton source size. 
One advantage of this ansatz is that
it facilitates comparison to interferometry radii. We show in Sec. III that 
this model
works 
well for our data since we find consistency between our estimate of the proton source size and Hanbury Brown-Twiss (HBT) results.  
At lower 
beam energies 
$R_{G}$ has been found to be very consistent with radii measured from $\pi\pi$, $KK$ and $pp$ correlations \cite{MurrayFaze}. 
The deuteron wave-function is more accurately represented by the Hulthen form, which has an exponential tail  \cite{HODGSON}.   
For the $B_2$ values reported in this paper the error in $R_G$ from using Eq.~\ref{eq:Rg}  
is less than 0.2 fm \cite{MurrayFaze}. 

For a variety of ion beams ranging from He to Ar incident on nuclear targets at BEVALAC energies \cite{COALDATA}, as well as $p$+A fixed target experiments at FNAL  \cite{Cronin:1974zm}, KEK \cite{Saito:1994tg} and SPS \cite{Bussiere:1980yq}, the measured $B_2$ values are independent of 
 energy
and $p_T$ and are consistent with measurements of the spatial parameters of the deuteron wave-function.  
At higher A+A energies including AGS Au+Au fixed target \snn=4.9 GeV \cite{Armstrong:2000gd,Armstrong:2000gz,Aoki:1992mb,Barrette:1997fj,E866,E896},  SPS  S+S, S+Pb and Pb+Pb at \snn $\approx$ 18 GeV \cite{NA44dp,physRev85,Anticic:2004yj}, and Au+Au (\snn=200 GeV) at the RHIC collider \cite{Adler:2004uy,Afanasiev:2007tv,StarDbarV2},  $B_2$ decreases with energy and increases with $p_T$.  This is consistent with the formation of deuterons 
 in an expanding medium  \cite{dasGupta,sato,Heinz}. 
Near midrapidity  $B_{2}$  decreases by a factor of $ \approx 20$ as the center of mass energy increases from  \snn $\approx$ 1 to 17.3 GeV  \cite{physRev85} 
before flattening out at RHIC energies \cite{STAR,PHENIX}.  
This is similar to the behavior seen in HBT radii \cite{Adler:2004rq, HBTvE}. 

\subsection{Average Phase-space Density}
The average of the phase-space density $f({\bf x},{\bf p})$ over the system volume at freeze-out time is a quantity, which, when compared to Bose-Einstein or Fermi-Dirac statistics, allows the assessment of the degree of chemical or kinetic equilibrium reached by the system at that stage \cite{Bertsch:1994qc,Pal:2003rz}. This quantity also  carries information about the possible multiple-particle symmetrization effects {\it i.e.} pion condensates. 
Values greater than 1.0 would  
indicate the presence of quantum effects \cite{Bertsch:1994qc}.  
This quantity is a count of the states that
the system can occupy, and as such, is used to define the contribution of particular types of particles to
the overall entropy of the system.
The spatial average of the phase-space density is defined as the following ratio:  

\begin{equation}
 \langle f \rangle ({\bf p}) \equiv \frac{\int d^{3}x f^{2}({\bf x},{\bf p})}{\int d^{3}x f({\bf x}, {\bf p})} ,
 \label{eq:fAverageDefine}
\end{equation}
where the integration is carried over spatial coordinates bound by the homogeneity volume of the system at freeze-out. The formal definition of phase-space density $ f({\bf x},{\bf p})$ for a particle of spin $J$ is written as:

\begin{equation}
 f( {\bf x}, {\bf p}) \equiv \frac{(2\pi \hbar)^3}{(2J+1)}
  \frac{d^6N}{dp^3dx^3} .
 \label{eq:fdefine}
\end{equation}

For a system in chemical  equilibrium at a temperature  $T$ and chemical potential $\mu$, 

\begin{equation}
 f(E) = \frac{1}{e^{(E-\mu)/T}\pm 1} ,
 \label{eq:fbose}
\end{equation}
where $E$ is the energy and $\pm 1$ selects bosons or fermions respectively.
For a dilute system, ({\it i.e.} $f \ll 1$),
Eq.~\ref{eq:fbose} gives:

\begin{equation}
 f_d  \approx  e^{-(E_d-\mu_p-\mu_n )/T} .
 \label{eq:fpnd}
 \end{equation}

 Since $E_d = m_T$cosh($y$), one would expect the phase-space density to be an exponential in $m_T$. 
 Note in this simple derivation,  we are ignoring the collective motion of the of particles.  
 At \snn~= 17.3 GeV, it was found that strong longitudinal flow could significantly reduce  
 the space-averaged phase-space density of pions \cite{Tomasik:2001uz}.   
Also at this energy 
the inverse slope of the phase-space density was found to increase with particle mass  in a manner suggestive of transverse flow \cite{Murray:2002ek}. 
Using the fact that the deuteron has energy $E_d = E_n + E_p$ and momentum ${\bf P} = 2{\bf p}$, Eq.~\ref{eq:fpnd} implies that

\begin{equation}
 f_d({\bf x}, {\bf P})=f_p({\bf x}, {\bf p}) \cdot f_n({\bf x}, {\bf p}) = f_p^2({\bf x}, {\bf p}) .
\end{equation}
To extract the average phase-space density of protons one can then replace the square term in the numerator of Eq.~\ref{eq:fAverageDefine}  by the phase-space density of deuterons:
\begin{equation}
 \langle f_p \rangle ({\bf p})= \frac{1}{3}(E_d\frac{d^3N_d}{dp_d^3}) / (E_p\frac{d^3N_p}{dp_p^3}).
 \label{eq:faze}
\end{equation}
Furthermore, one  can make use of the assumption that deuterons are formed by coalescence and satisfy Eqs.~\ref{eq:coal-eq} and \ref{eq:Rg} to obtain an expression for the average phase-space of protons similar to what Bertsch originally suggested for pions. Dividing the spectrum by the product of the HBT radii gives \cite{Bertsch:1994qc, Pal:2003rz}:
\begin{equation}
 \langle f_p \rangle({\bf p}) = \frac{1}{2}  \cdot E_p\frac{d^3N_p}{dp^3}\frac{{\pi}^{3/2}\cdot{\hbar}^3}{R_G^3({\bf p}) \cdot m_p c^2} .
  \label{eq:fazeRg}
\end{equation}

The phase-space densities calculated using Eq.~\ref{eq:faze} have the expected exponential dependence in $m_T$. When using Eq.~\ref{eq:fazeRg} one introduces the assumption that deuterons are produced via coalescence and that the homogeneity volume extracted from deuteron distributions is also the volume of the proton source. The calculation of the space-averaged phase-space density 
of protons using these assumptions is close in value to the one obtained from an assumed thermal equilibrium and has the same behavior in $m_T$. This suggests that the system has evolved into chemical equilibrium at freeze-out.  While chemical equilibrium has been very well established at RHIC energies for the system at midrapidity, these data allow us to test if the matter at forward rapidity also reaches equilibrium.  
To ease the comparison with lower energy data we have decided to use Eq.~\ref{eq:faze} to calculate the phase density.

The coalescence parameter $B_{2}$ and the space-averaged phase-space density recast the information contained in the proton and deuteron spectra into ``dynamic" and ``chemical" terms.
The coalescence parameter $B_2$ can be interpreted in terms of the homogeneity volume, which depends upon the temperature of the system and the radial flow.  
 Indeed one confirmation that we are actually seeing coalescence would be to see if  this radius was consistent with the appropriate HBT radii. This was checked at \snn=17.3~GeV by comparing $R_G$ to radii  extracted from $\pi\pi$, $KK$, and $pp$ correlation measurements 
\cite{MurrayFaze,NA44ppHBT,Beker:1994qv,Bearden:2001sy}.   
In this paper, we use the very large angular and momentum ranges of the two BRAHMS spectrometers to  measure the rapidity dependence of the homogeneity volume  and space-averaged phase-space density of the (anti-)proton distributions for central  Au+Au collisions.

\section{Analysis}\label{analysis}
The data in this paper were collected by the BRAHMS experiment during  2004. We present proton and deuteron spectra at \snn~=200 GeV  AuAu collisions with a centrality range of 0-20\%. 
The statistics for deuterons at high rapidity for more peripheral data is limited, so a centrality dependence analysis was not warranted.
The data are analyzed 
 in four rapidity bins: $(-0.1,0.1)$, $(0.5,1.2)$, $(1.5,2.5)$, and $(2.8,3.2)$.  
We have also measured antiproton and antideuteron spectra at $y\approx 0$ and $y\approx 0.8$.

\subsection{\label{sec:detector}Detector System}
The BRAHMS experimental setup consists of two movable magnetic
spectrometers,  the Forward Spectrometer
(FS) that can be rotated from \degree{2.3} to \degree{15}, and the
mid-rapidity Spectrometer (MRS)
that can be rotated from \degree{90}\  to \degree{30}\  degrees relative to the beam line.
Fast Cherenkov counters at high rapidities were used to
measure luminosity, to determine the interaction vertex, and
to provide a start time for time-of-flight measurements.

The MRS has two time projection chambers (TPCs), TPM1
and TPM2, situated in field free regions in front and behind a dipole magnet.
This assembly is followed by a highly segmented scintillator time-of-flight (TOF)
wall at 4.51 m. 
The FS consists of 4 dipole magnets D1, D2, D3 and
D4 with a total bending power of up to 9.2 Tm. 
The spectrometer has 5 tracking stations T1 through T5.
T1 and T2 are TPCs placed in front of 
and after the second dipole D2. T3, T4 and T5 are drift chambers with T3 in front of
D3, T4 between D3 and D4, and T5 after D4 and just in front of the
 ring imaging Cherenkov (RICH)~\cite{RICH}.  Details on the BRAHMS experimental setup can be found in
Ref.~\cite{BRAHMS:NIM}.

\subsection{Centrality selection}
The centrality of the collisions at the BRAHMS interaction region was extracted from charged particle multiplicity densities measured with a  multiplicity array (MA).
The MA array consists of a coaxial arrangement of
Si strip detectors and scintillator tiles 
surrounding the intersection region \cite{Lee:2004su}.
The pseudo-rapidity coverage of the MA is approximately 
$-2.2<\eta<2.2$.
The centrality selection is derived from 
minimum-bias trigger events, which are defined 
using two zero degree calorimeters (ZDC), 
requiring energy deposit equivalent to at least
one neutron in each of the two detectors
and also requiring a signal in the MA in order 
to reject Coulomb dissociation events~\cite{200gevmult,BRAHMS:NIM}.
The uncertainty in the centrality determination from 
our multiplicity distributions 
is estimated to be 
$\pm 4\%$ of the size of the bin for the 0\%-20\% bin.
The fraction of the inclusive yield lost by the minimum-bias trigger is 
estimated to be about 4\% and was corrected for.

\subsection{Tracking}

Track reconstruction starts by finding local track segments in the TPCs and (for the forward 
spectrometer) in the drift chambers. These detectors are in field free regions and so these local 
tracks are straight lines. Local tracks on either side of a magnet are matched using
 the effective edge approximation generating what we call
locally matched tracks. 
The locally matched tracks are then combined in the FS to form complete tracks. 
The complete tracks are refitted to extract the best measurement of the particle three momentum.
Tracks
in the FS are required to project through the magnet D1 onto the
nominal beamline. 
Track quality cuts are applied for the final track selection.
The momentum resolution at full field ( where all data presented in this paper were taken) is: $\delta p / p = 0.0008 p$. 

The  momentum distributions  are corrected for tracking detector efficiency and geometrical
acceptance. 
Tracking and matching efficiencies for each of the 5 tracking stations in 
the FS
 were calculated by constructing full tracks using just
four track segments and evaluating the efficiency in the fifth 
station by comparing the predicted position and direction of the interpolated or extra-polated 
full track in that station with the known local segments.
The local track efficiency was evaluated as function of spectrometer angle and field setting, as well as position and direction of the local track segments.  
The overall tracking efficiency is typically 60\%, and is known to a few percent.  
In the MRS the tracking efficiencies are determined from Monte Carlo simulation using embedding techniques. The systematic error is below $4\%$.
We correct for absorption, multiple scattering,  
and energy loss in the detectors using GEANT 3.21 \cite{GEANT}. 
The magnitude of these corrections on the particle yields depends on the 
particle momenta and the spectrometer positions, and in the FS it is about 
 $20\%$ with an estimated uncertainty of $2\%$. In the MRS the corrections are less than $8\%$ with 
a small systematic error.
We assume that deuteron absorption factorizes from other effects and model 
it as the square of the  proton absorption correction (at $p_T/2$) as done 
in \cite{MurrayFaze}.  More details about tracking in the MRS
can be found in Ref.~\cite{BRAHMS:auau2005PRC}.

\subsection{Particle Identification}
\begin{figure}[htb]
   \includegraphics[angle=90,width=\columnwidth]{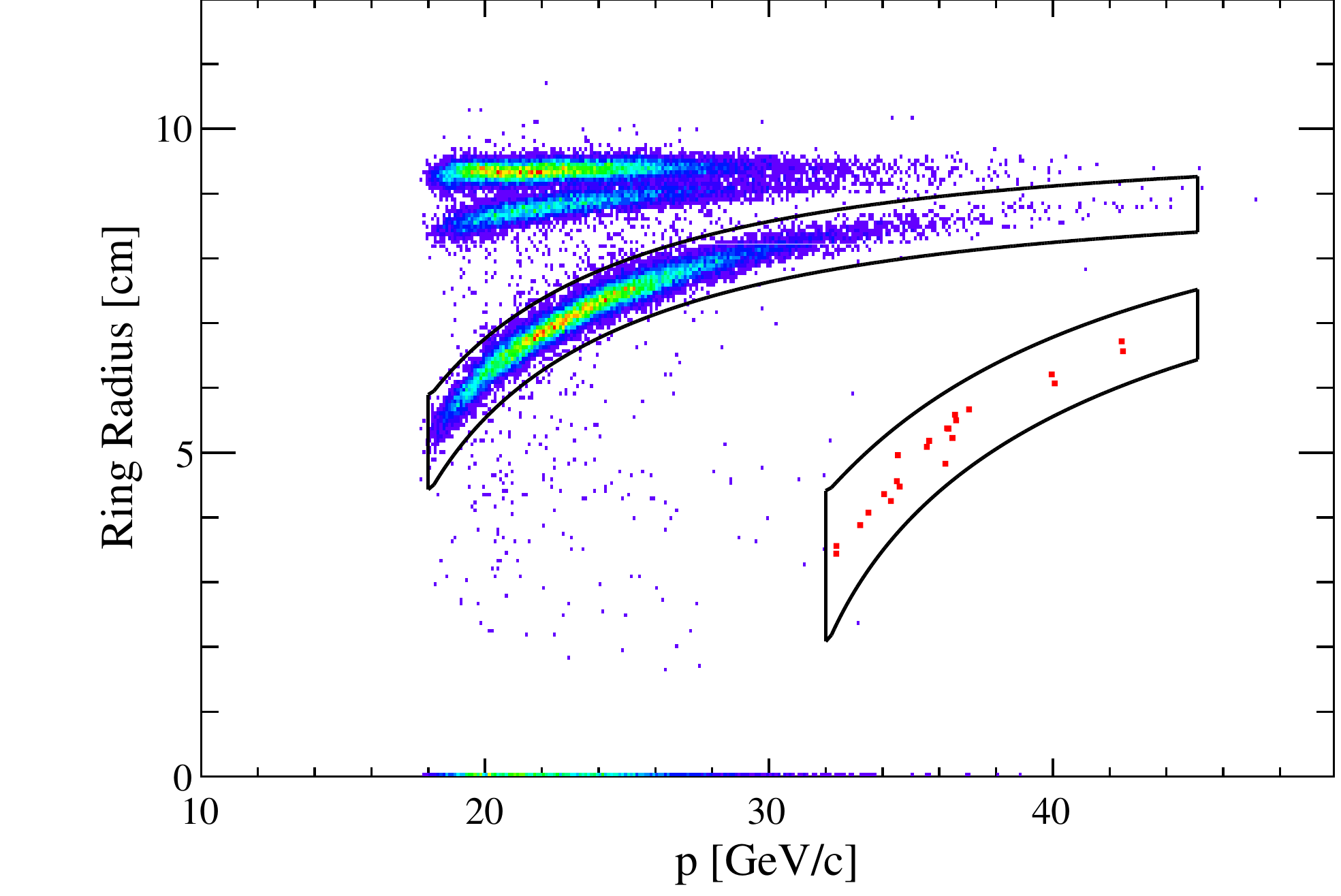}
  \caption{\label{Fig:PID} (color online) Ring radius versus momentum for particles at $y \approx 3$ showing the pion, kaon, proton and deuteron separation. 
  The bands show the proton and deuteron selection  used in the analysis. }
\end{figure}

In the midrapidity region, particle identification  is done using
time of flight measurements, while at forward rapidities we use the RICH 
~\cite{RICH}.  At $y$=0 and 0.8 we select 
(anti)protons using a 3$\sigma$ cut on $m^2= p^{2}  (\frac{c^2L^2}{t^2} - 1)$, where $L$ is the length of the path followed by a particle, $t$ its time of flight, and $c$ the speed of light. For deuterons a simple  3$\sigma$ cut around the (anti)deuteron mean mass is used.
These deuteron yields are corrected for background that arises from false matching of tracks with background hits in the TOF wall. This background is determined from data in the mass region outside the deuteron peak, and contributes to the systematic error of the deuteron yields by $\sim 5-10\%$ decreasing with \pt.
At forward rapidities the RICH provides proton/deuteron 
PID separation.
The ring radius ($R$) in the RICH depends on the particle's velocity.  For a particle of mass $m$ in the RICH
$R \approx 9.2 (1-(p/p_{thres})^2)$ cm where $p_{thres} =  16.1\cdot mc$. 
The PID performance is shown in Fig.~\ref{Fig:PID}, where the bands represent the limits used for the analysis.   
Because of inefficiencies about 3\% of  the particles  moving faster than their Cherenkov
threshold don't produce an identified ring.  
Protons and deuterons with momenta between the kaon and proton thresholds do not make a ring. 
The contamination from mis-identified kaons and pions is subtracted to deduce the proton spectrum in this momentum range (neglecting the small fraction of deuterons). 
The contamination correction to the proton spectrum is $\sim10\%$ at 
\pt=0.5 \gevc~ and drops to $\sim 2\%$ at \pt=1.5 \gevc. The error on this 
correction is neglible.  
This procedure cannot be used in the region between the proton and deuteron thresholds since the contamination from pions, kaons, and protons dominates the small deuteron yield.

\subsection{Feed-down corrections}
We have corrected our data to account for the hyperons that  decay into
protons using the method described in \cite{BRAHMSstopping}.    
The correction factor, $C$, is given by:
\begin{equation}
C=\frac{N_p}{N_p +  N_{\Lambda}+  N_{\Sigma^+}} ,
\end{equation}
where $N_p$ is the number of primary protons and $N_{\Lambda}$ and and $N_{\Sigma^+}$ are the
number of protons coming from $\Lambda$ and $\Sigma^+$ decay, respectively. 
Near  central rapidity we have used the $\Lambda$ spectra measured by PHENIX \cite{Adcox:2002au} and estimated the $\Sigma/\Lambda$ ratio from lower energy measurements \cite{BRAHMSstopping}. 
Since there are no measurements of $\Lambda$s at forward rapidities we have estimated the $\Lambda/p$ ratio based on thermal models that were fitted to the rapidity densities of charged pions, kaons, protons and antiprotons measured by BRAHMS in the forward region 
\cite{Stiles:2006sa}.

\begin{figure}
\includegraphics[height=\columnwidth, angle=90] 
{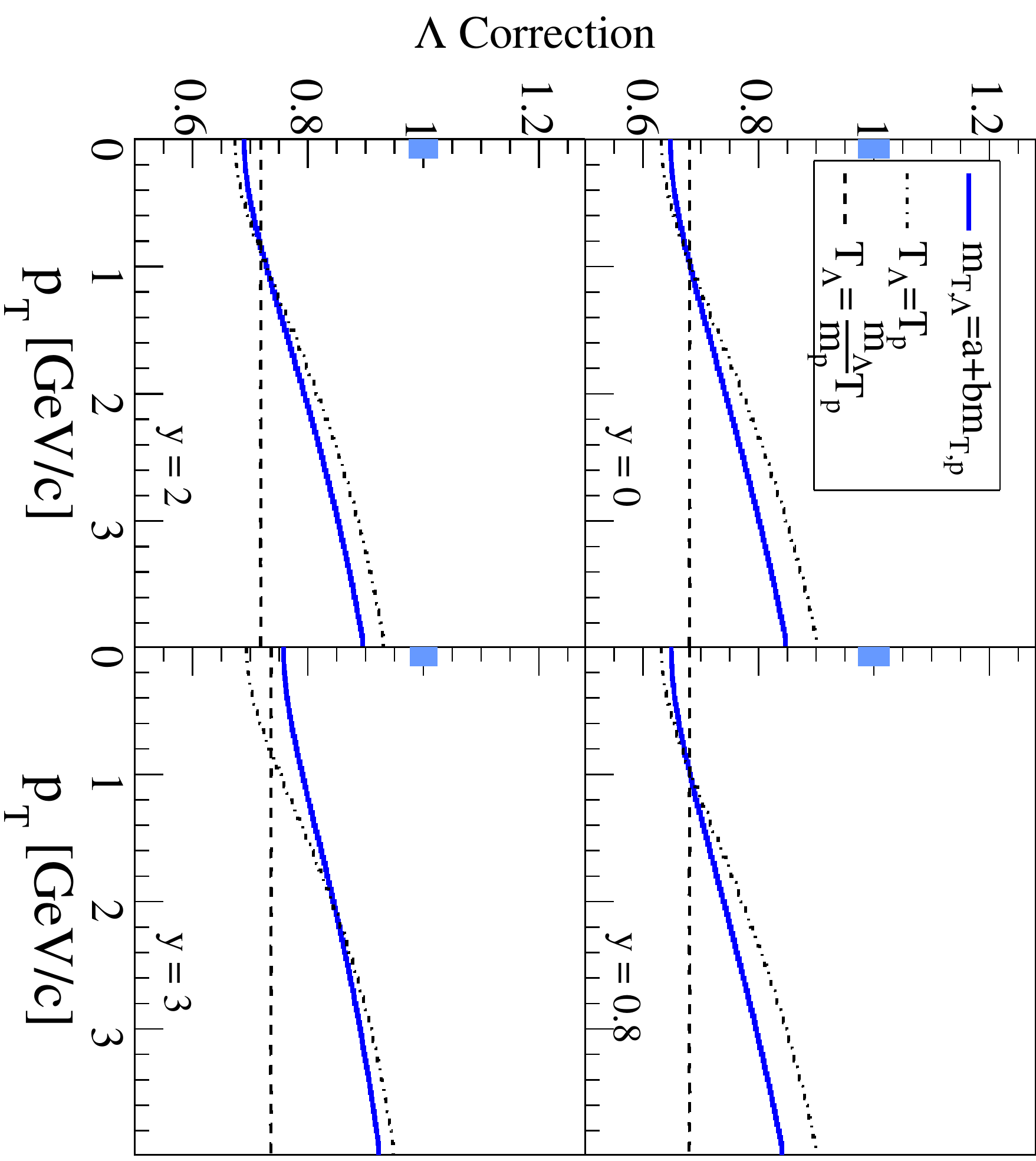}
\caption{\label{Feeddown} (Color online) \pt\ dependent feed-down correction factors for protons versus \pt~ and rapidity.  
The three sets of lines show the effect of various scenarios for the difference in shape  between proton and $\Lambda$ spectra. 
Systematic errors on the correction due to uncertainties in the yield ratio of protons and lambdas are shown by the bands to the left of the plots. The curves for antiprotons are very similar.  } 
\end{figure}

With our model calculations, we find that the ratio $\frac{dN_\Lambda}{dy}/\frac{dN_p}{dy}$ varies
 slowly with rapidity up to rapidities $y\approx 4$.
The systematic error from uncertainties on the yields and the model extrapolation is estimated to be less than 3\%. The correction factor also depends on the $p_T$ dependence of the $\Lambda/p$ ratio. 
BRAHMS has found that the mean transverse kinetic energy  scales linearly with the  mass of the particle with a slope that depends only weakly on rapidity \cite{SandersQM09}. 
We have used these systematics to estimate the inverse slope ($T$) of the $\Lambda$  $m_T$-distribution.  
To estimate the systematic error on the $p_T$ dependence of the correction factor we have taken the limiting cases of $T_{\Lambda} = T_p$  and $T_{\Lambda} = T_p \cdot m_{\Lambda}/m_p $.   This produces an error on the correction factor that is almost zero at 
$p_T$ = 1~\gevc\ and reaches $-9$\%, +6\% 
at $p_T$ = 2~\gevc.   The correction factor as a function of $p_T$ and rapidity is shown in Fig.~\ref{Feeddown}.  It varies only weakly with rapidity and has a systematic error that is small in comparison to the statistical errors on $B_2$  and the space-averaged phase-space density.  
At $p_T$ = 2~\gevc\ the total error from the feed-down correction reaches a value of 19\% for $B_2$  and 10\% for the space-averaged phase-space density. 

\section{Results}\label{results}
\begin{figure*}
\includegraphics[height=\textwidth, angle=90]
{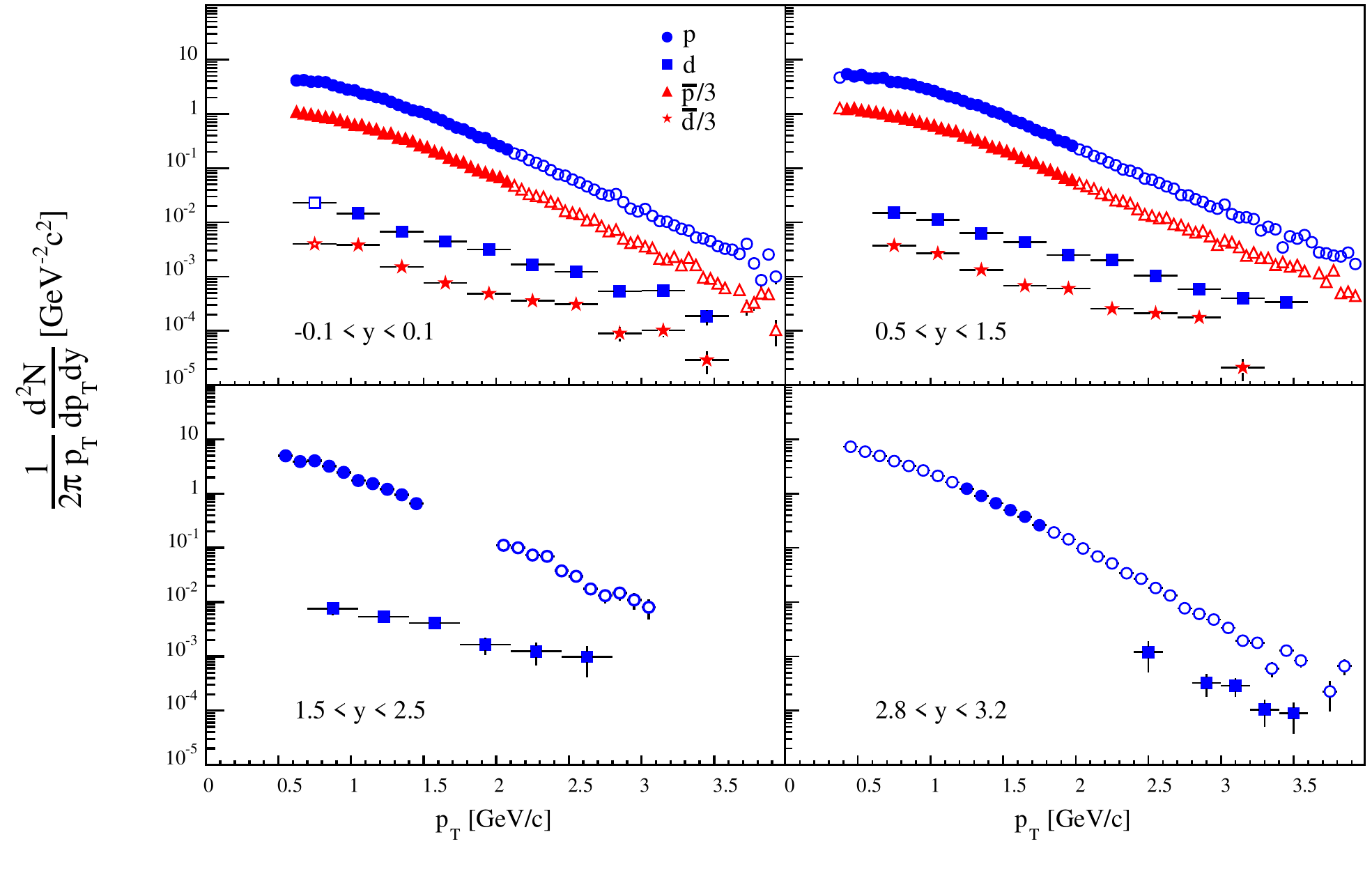}
\caption{\label{spectra} (color online) (anti-)proton and (anti-)deuteron 
\pt~spectra at various rapidities for the top $20\%$ most central Au+Au collisions.  The filled symbol part of the spectra  show the \pt~ intervals used in the coalescence analysis. Note the limits for the deuteron range are exactly twice those used for the protons.
The errors are statistical only. The horizontal bars represent the bin width.} 

\end{figure*}

Figure \ref{spectra}  shows the invariant proton and deuteron spectra versus $p_T$ for the four measured rapidity bins.
The systematic error on the normalization of the spectra is estimated
to be about $5\%$ with a point-to-point error in the spectra  of $\approx 8\%$ arising from 
the merging of spectrometer settings and PID and uncertainties in tracking efficiency determination.
We fit exponentials in $m_T$ to the spectra and extract the invariant yields, $dN/dy$  and inverse slopes $T$.  
These are listed in Table \ref{tab:results}. 
About  40-50\% (dependent on rapidity)
of the yield of the deuterons is contained within our acceptance. For protons the fractions are slightly higher 55-60\%. 
The proton yield falls slightly from y=0 to y=3, but 
 no firm conclusion can be made for the deuterons due to the limited $p_T$ range at high rapidity. 
  The inverse slopes tend to decrease with rapidity, suggesting a decrease of radial flow.  The inverse slopes of the deuterons are somewhat higher than those of the protons. The average over all rapidities for the ratio of the inverse slopes  is $1.6 \pm 0.1$.  For antideuterons the inverse slopes are a factor of $2.0 \pm 0.2$ higher than that of antiprotons. 

\begin{table}
\begin{center}
\begin{tabular}{|l|c|l@{$\,\pm\,$}l|l@{$\,\pm\,$}l|c|l@{$\,\pm\,$}l|l@{$\,\pm\,$}l|}
\hline
    & \multicolumn{5}{c|}{ Proton} & \multicolumn{5}{c|}{ Deuteron}  \\ 
$y$      & $p_T$ fit  & \multicolumn{2}{c|}{$dN/dy$} & \multicolumn{2}{c|}{$T$ MeV }  &     $p_T$ fit     & \multicolumn{2}{c|}{ $dN/dy$} & 
\multicolumn{2}{c|}{ $T$ MeV } \\ \hline
\hline
0.0 & 0.7-4.0    & 27.9  & 0.1 & 354  & 2   & 1.5-3.3  & 0.093 & 0.008      & 570  & 70  \\ 
       &               & 20.8  & 0.1  & 352 & 1   &               &  0.033 & 0.004     & 870 & 220  \\  \hline      
0.8 &  0.5-4.0   & 26.0  & 0.1  & 356&1    & 1.5-3.3 &  0.068 & 0.003       & 610 &30  \\
        &              & 17.9  & 0.1   & 361 &1   &             &  0.031 & 0.002       & 700 &80 \\  \hline
 2.0  & 0.5-3.0  &  20.9  & 0.2  & 314 &3    & 0.8-2.8  & 0.082  & 0.011      & 460 &110 \\ \hline
 3.0  & 0.5-4.0  & 23.4  & 0.1   & 282 &1    & 2.5-3.5   &  \multicolumn{2}{c|}{$0.25^{+0.30}_{-0.08}$}     &310 &90   \\  \hline

\hline 
\end{tabular}
\end{center}
\caption{\label{tab:results} Proton and deuteron yields, $dN/dy$, and inverse slopes, $T$ (MeV),    derived from fitting spectra. The lower rows at  rapidities  $y=0$ and $y=0.8$ are for the anti-particles.  The errors listed are statistical only. For $dN/dy$ the systematic errors due to extrapolation to low pt is 5-10\% with an additional 8\% from normalization  and various corrections.  The systematic errors on the inverse slope are less than 10\%. }
\end{table}

\begin{figure}
\begin{center}
\includegraphics[height=\columnwidth, angle=90]{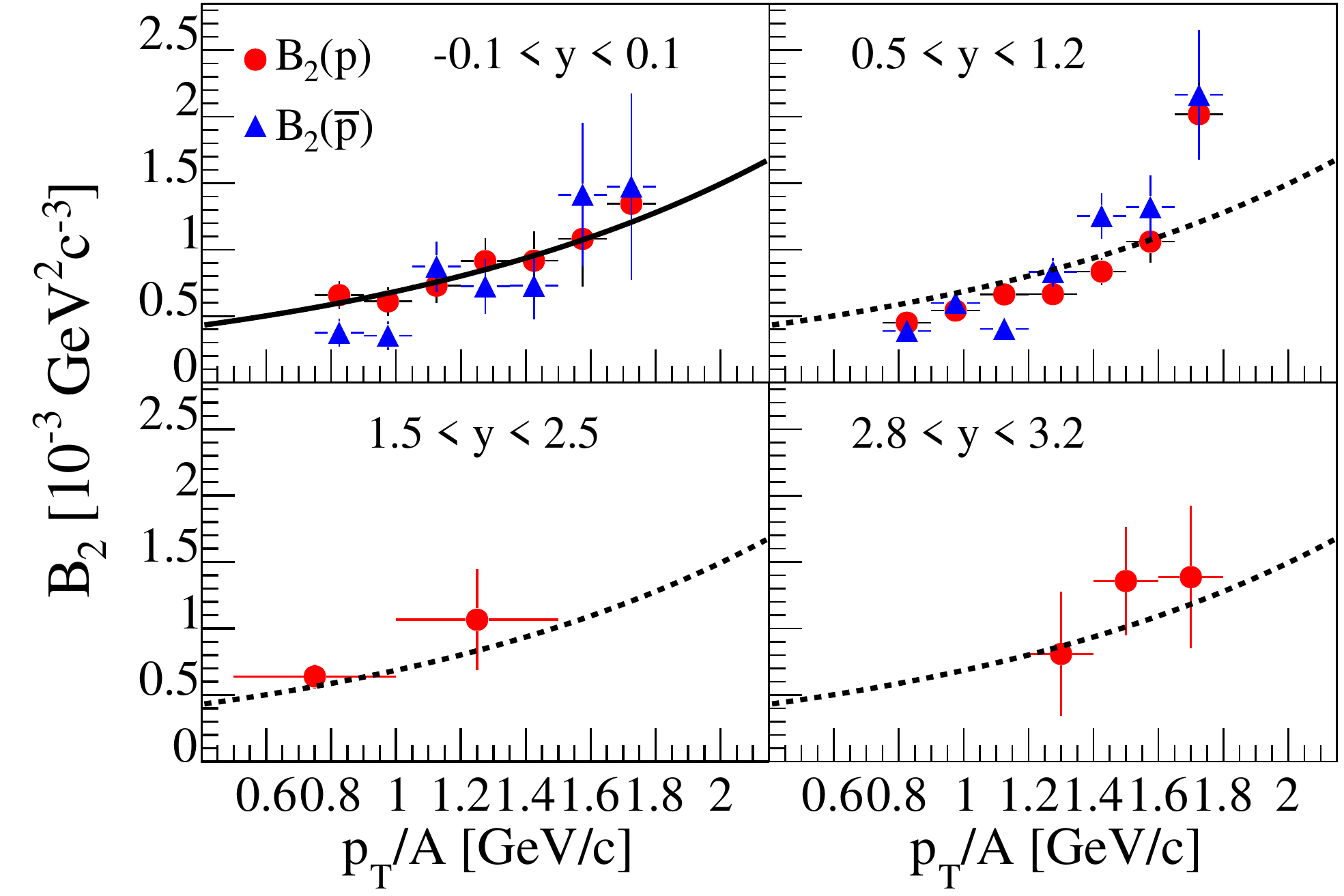}
\caption{\label{b2vspt}  (color online) $B_{2}$ versus transverse momentum per nucleon at several rapidities for central Au+Au collisions at \snn~= 200 GeV. Protons are displayed with filled circles and antiprotons with filled triangles. The solid line in the top left panel is an exponential fit to the data at y=0. This same line is shown, in dotted form, in the other 3 panels. 
The errors are statistical only.}
\end{center}
\end{figure}

Figure \ref{b2vspt} shows $B_2$ versus $p_T$ and rapidity.  
$B_2$ increases with $p_T$, 
 which is consistent with previous experiments \cite{physRev85,PHENIX}.  
 Using Eq.~\ref{eq:Rg}, we find that at central rapidity the source radius $R_G$ falls from $4.2 \pm  0.2 $ fm to $3.1 \pm 0.4$ fm as $m_T$ increases from 1.2 to 1.9 ${\rm GeV}/c^2$. 
 This is consistent with the $m_T$ dependence of  HBT radii, $R \propto 1/\sqrt{m_T}$ that has been observed by  PHENIX \cite{Adler:2004rq} and STAR \cite{HBTvE} and also at SPS energies \cite{Beker:1994qv}.  
 The solid line in each rapidity panel represents an exponential fit to our data at $y$=0. We see no evidence of any rapidity dependence of $B_2(p_T). $ The proton and antiproton $B_2$ values are very close at this energy implying a similar source size. This is in contrast to the measurements at \snn~= 17.3 GeV where the antiproton source volume was found to be somewhat larger than the proton source volume \cite{MurrayFaze}. 

\begin{figure}
\begin{center}
\includegraphics[height=\columnwidth, angle=90]
{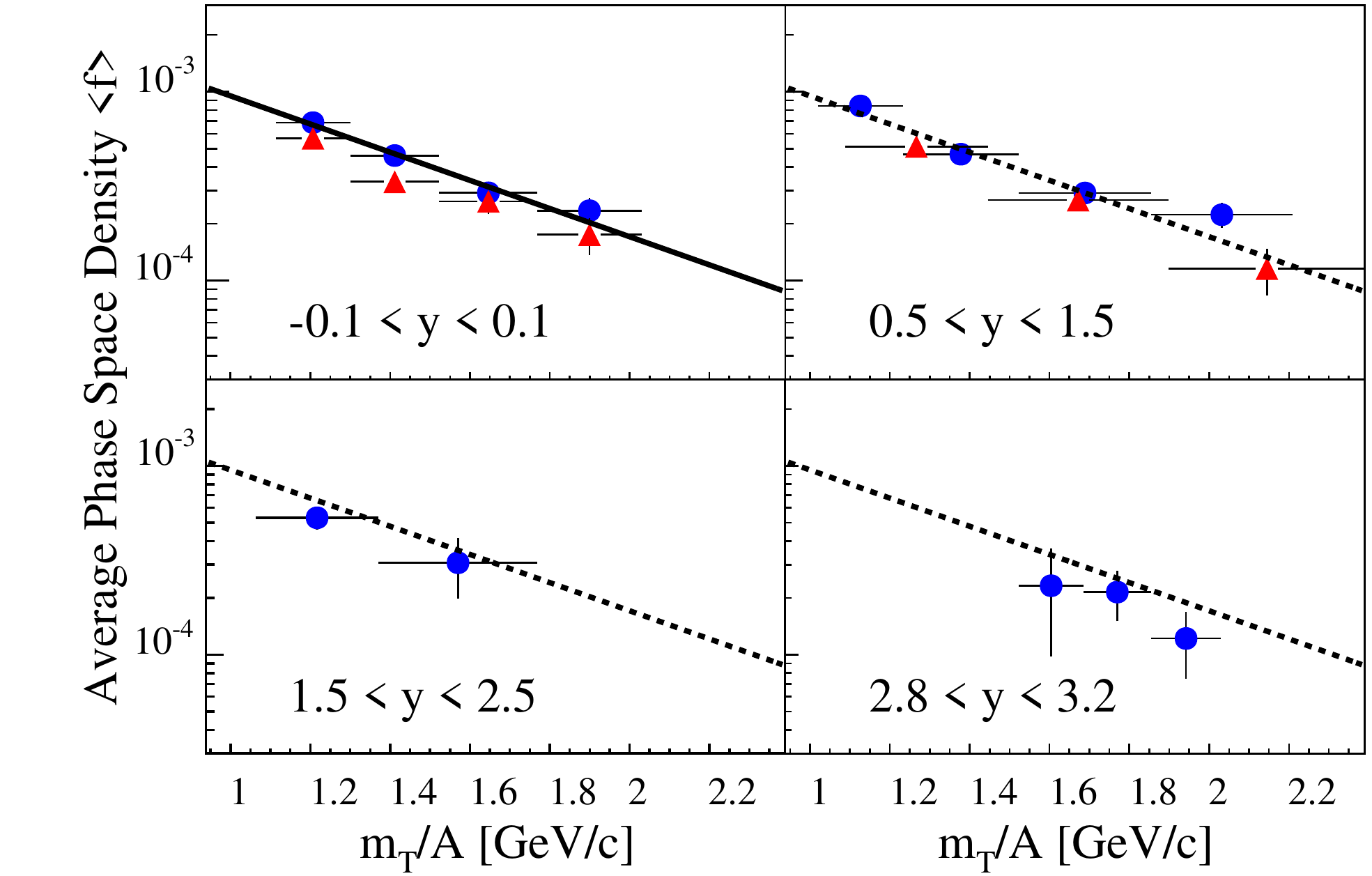}
\end{center}
\caption{\label{fvsy} (color online) The  (anti-)proton space-averaged phase-space density $\langle f \rangle (m_{T}/A)$ for central \snn~ = 200 GeV Au+Au collisions at several 
rapidities.  The solid line in the top left panel is an exponential fit to the data at y=0. This same line is shown, in dotted form, in the other 3 panels. 
The errors are statistical only.}
\end{figure}

Figure ~\ref{fvsy} shows the average phase-space density $\langle f \rangle (m_T)$ for protons and antiprotons as a function of rapidity. 
These values are calculated using Eq.~\ref{eq:faze}.  
The space-averaged phase-space density decreases as the  $m_T$ increases 
as expected from Eq.~\ref{eq:fbose}. The solid curve in each panel of Fig.~\ref{fvsy} is an exponential fit to the proton density at $y$=0. We see little rapidity dependence of $\langle f \rangle (p_T)$.   

\begin{figure}
\begin{center}
\includegraphics[height=\columnwidth]
{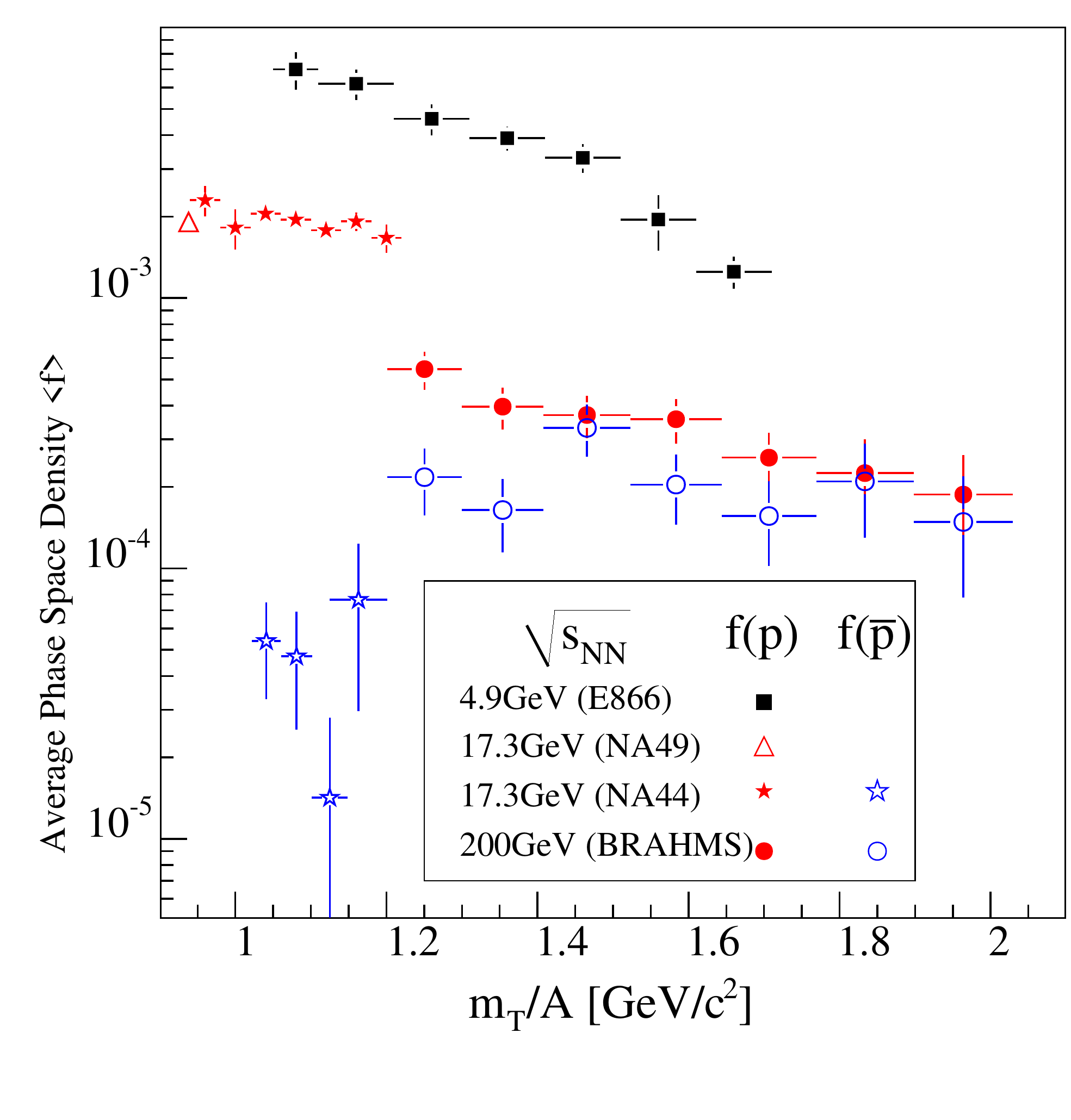}
\end{center}
\caption{(color online) The proton space-averaged phase-space density for central collisions as a function of \snn~and  $m_T$ at $y$=0 \cite{E866,Anticic:2004yj,MurrayFaze, physRev85}.
The errors are statistical only.
}\label{fvsE}
\end{figure}

  From  Eq.~\ref{eq:fbose} we would expect the ratio of the proton and antiproton phase densities to be flat as a function of $p_T$.  Fitting a constant to $\langle f_{\bar p} \rangle/\langle f_p \rangle$ yields a $\chi^2/NDF$ of 5.0/6 and 13.3/6 at $y=0$ and $y=0.8$, respectively.  
  Near $y$=0 the antiprotons have a slightly smaller value of space-averaged phase-space density compared to that of protons, suggesting a small chemical potential. At  \snn~= 17.3 GeV the antiproton space-averaged phase-space density was 38  times smaller \cite{MurrayFaze}, which suggests a much larger baryo-chemical potential at the lower energy.  The inverse slope derived from the space-averaged phase-space density is  $T = 930 \pm 110$ MeV for protons. This is consistent with data at \snn~= 17.3~GeV, but much higher than the value extracted at  \snn~= 4.9 GeV where the inverse slope is  about 350 MeV, (see Fig.~\ref{fvsE}).  It should be noted that the proton, kaon and pion spectra 
at midrapidity can be well described by blast-wave fits,  which suggest that this increase in the inverse slope with \snn~  is largely driven by an increase in radial flow.   
This is supported by the fact that the phase-density of pions is characterized by a much smaller inverse slope ($\approx 140$ MeV) than that of protons \cite{MurrayFaze}.   

Figures~\ref{b2vspt} and ~\ref{fvsy} imply that the volume of homogeniety, $\sim 1/B_2$, and the space-averaged phase-space density vary little  with rapidity at \snn~ = 200 GeV. This is in stark contrast to the situation at lower energy.  
Table ~\ref{tab:17GeV} shows the rapidity dependence of $B_2$ and the space-averaged phase-space density at $p_T = 0$ for central Pb+Pb  collisions at \snn~= 17.3~GeV \cite{Anticic:2004yj,NA44dp,MurrayFaze}.  These data are compared to BRAHMS results at $p_T/A$ = 1.3 GeV/$c$ which show very little rapidity dependence. 
Finally, Fig.~\ref{fvsE} shows the evolution of the space-averaged phase-space density near midrapidity as the energy of the system grows from AGS to RHIC values. The 
space-averaged phase-space density of protons decreases with energy while that of antiprotons increases.  
At AGS energies $\langle f \rangle$ at $m_{T}=0$ is 10 times bigger than its value at RHIC, while at
SPS it is 2 times bigger. In contrast, the pion average phase-space density has the opposite behavior as the energy
of the colliding system increases; the values at RHIC \cite{Pal:2003rz} are higher than those at SPS by a factor 
$\sim 2$ 
where $\langle f \rangle=0.45$ \cite{Ferenc}.  At \snn 4.9 GeV $\langle f \rangle$ was extracted from  HBT and spectra measured
at high rapidity $\langle y \rangle \sim 3.1$ in Au+Au 10\% central collisions and its estimated value at midrapidity
is $\sim 5$ times smaller that the RHIC value \cite{Barrette:1997fj}. For antiprotons the 
growth in $\langle f \rangle$ is similar to that of pions.  
The different trend for the protons may be driven mainly by baryon transport; at AGS, most of the beam protons end 
up at midrapidity, less so at SPS, and at RHIC, less than 10\% of the beam protons are transported to midrapidity.
 The proton rapidity loss adds beam protons to the Gaussian
rapidity density of produced protons and transforms the overall proton rapidity density into a flat distribution.

\begin{table}[ht]
\begin{center}
\begin{tabular}{|c|l@{$\,\pm\,$}l|l@{$\,\pm\,$}l|l@{$\,\pm\,$}l|l}
\hline
$(p_T/A = 0)$      &   \multicolumn{2}{c|}{ $y$=0.2 } & \multicolumn{2}{c|}{ $y$=0.8}     & \multicolumn{2}{c|}{$y$=1.3}  \\ \hline
$B_2 \times 10^4 $ & 7.9 &  0.8 &  8.1 &  0.3 &  13.7 & 2.7 \\ \hline
$f \times 10^3$          & 1.9 & 0.1 & 2.5 & 0.2 & 3.3 & 0.3 \\ \hline
\hline 
\end{tabular}
\protect{\vspace{0.3cm}}
\begin{tabular}{|c|l@{$\,\pm\,$}l|l@{$\,\pm\,$}l|l@{$\,\pm\,$}l|l@{$\,\pm\,$}l|l}
\hline
$(p_T/A = $1.3 GeV/c)      &   \multicolumn{2}{c|}{ $y$=0.0 } & \multicolumn{2}{c|}{ $y$=0.8}     & \multicolumn{2}{c|}{$y$=2.0}& \multicolumn{2}{c|}{$y$=3.0}  \\ \hline
$B_2 \times 10^4$ & 8.1 & 1.0 & 6.6 &  0.6 &  10.6 &   0.4 & 8.1 & 0.5 \\ \hline
$f  \times10^4$      & 3.6 & 0.6 & 2.4  & 0.3 &    3.1   & 1.0 & 2.3 & 1.0 \\ \hline
\hline 
\end{tabular}
\end{center}
\caption{\label{tab:17GeV}  The rapidity dependence of  \protect{$B_2$} and the $f$ for  (Top) Pb+Pb  collisions at \snn~= 17.3 GeV and (Bottom)  \snn~= 200 GeV AuAu collisions. For the 17.3 GeV data the centrality is 23\% at $y$=0.2 \cite{Anticic:2004yj} and 20\%  at $y$=0.8 and 1.3 \cite{NA44dp,MurrayFaze}.}
\end{table}

\section{Summary}
The rapidity dependence of the deuteron production in $0-20\%$ central Au+Au collisions at 200 GeV has been studied in the context of coalescence models.
Near central rapidity the proton and antiproton phase-space densities are very similar, suggesting a small baryon chemical potential. The coalescence parameters, or $B_2$-values,  for deuterons and antideuterons are also very close, 
suggesting similar freeze-out volumes for protons and antiprotons.  
$B_2$ increases with $p_T$ as expected for a system undergoing transverse flow; flow introduces a correlation between 
position and momentum.  At a given $p_T$ the deuteron coalescence parameter $B_{2}$ is 
independent of rapidity, which would imply that the homogeneity volume for protons of a given $p_T$ is almost constant from y=0 to y=3. Baryon transport in these colliding systems may be affecting the rapidity density of protons.  
 It is interesting to note that the radial flow  varies weakly with rapidity \cite{SandersQM09}. The weak dependence of radial flow on rapidity offers an explanation of why the $p_T$ dependence of $B_2$ does not depend on rapidity. 
It does not however address the fact that the overall magnitude of $B_2$ is constant from y=0 to y=3. 
The proton space-averaged phase-space density also shows no significant rapidity dependence, whereas it depends strongly on $m_T$ and \snn.  The space-averaged phase-space density can be thought of as the ratio of the number of protons per unity rapidity over the volume of the proton source.  The rapidity density of 
protons  decreases by only a factor of $0.84 \pm 0.01$ from y=0 to y=3. The constancy of the space-averaged phase-space density in rapidity is consistent with the similar behavior found for $ B_2$. 
Thus we observe a striking invariance of the proton source over several units of rapidity.

\vspace{0.5cm}

\section{Acknowledgements}
This work was supported by the Division of Nuclear Physics of the
Office of Science of the U.S. Department of Energy under contracts
DE--AC02--98--CH10886, DE--FG03--93--ER40773, DE--FG03--96--ER40981, and
DE--FG02--99--ER41121, the Danish Natural Science Research Council,
the Research Council of Norway, the Polish Ministry of Science and 
Higher Education (Contract no. 1248/B/H03/2009/36), and the Romanian
Ministry of Education and Research grant no. 81-049/2007 (REEHUC). We
thank the staff of the Collider-Accelerator Division at BNL for
their excellent and dedicated work to deploy RHIC and their
support to the experiment.

\end{document}